\documentclass[reprint, amsmath,amssymb, aps,prl]{revtex4-1}

\usepackage{graphicx}
\usepackage{dcolumn}
\usepackage{bm}
\usepackage{color}

\begin{document}

\title{Multi-modal excitation to model the Quasi-Biennial Oscillation}

\author{P. L\'eard$^{1,\dag}$}
\author{D. Lecoanet$^2$}
\author{M. Le Bars$^1$}

\affiliation{
 $^1$ Aix Marseille Universit\'e, CNRS, Centrale Marseille, IRPHE, Marseille, France\\
 $^2$  Northwestern University, Engineering Sciences and Applied Mathematics, Evanston, IL, 60208\\
 $^\dag$ corresponding author: leard@irphe.univ-mrs.fr
}

\date{\today}

\begin{abstract}
The Quasi-Biennial Oscillation (QBO) of stratospheric winds is the most striking example of mean-flow generation and reversal by the non-linear interactions of internal waves. Previous studies have used an idealized monochromatic forcing to investigate the QBO. Here we instead force a more realistic continuous wave spectrum. Unexpectedly, spreading the wave energy across a wide frequency range leads to more regular oscillations. We also find that different forcing spectra can yield the same QBO. Multi-modal wave forcing is thus essential for understanding wave—mean-flow interactions in nature.
\end{abstract}

\maketitle

Internal gravity waves (IGWs) are ubiquitous in geophysical and astrophysical flows, \textit{e.g.} in Earth's oceans \cite{garrett_internal_1979}, atmosphere \cite{fritts_gravity_2003,miller_upper_2015} and core \cite{buffett_evidence_2016}, as well as in stellar interiors \cite{charbonnel_mixing_2007,straus_energy_2008,rogers_internal_2013}. IGWs extract momentum from where they are excited and transport it to where they are damped \cite{bretherton_momentum_1969}. In the Earth's stratosphere, these waves drive oscillations of zonal winds at equatorial latitudes, with a period of nearly 28 months. This phenomenon, known as Quasi-Biennial Oscillation (QBO), affects, e.g., hurricane activity in the Atlantic ocean \cite{baldwin_quasi-biennial_2001}, and the winter climate in Europe \cite{marshall_impact_2009}. Similar reversals are observed on other planets \cite{fouchet_equatorial_2008,leovy_quasiquadrennial_1991}. The QBO is a striking example of order spontaneously emerging from a chaotic system \cite{couston_order_2018}, similar to magnetic field reversals in dynamo experiments \cite{berhanu_magnetic_2007} or mean-flow reversals in Rayleigh-B\'enard convection \cite{araujo2005wind}.

Atmospheric waves are excited by turbulent motions in the troposphere and propagate in the stratosphere, leading to zonal wind reversals.
Lindzen and Holton \cite{richard_s._lindzen_theory_1968,lindzen_updated_1972} proposed the QBO is due to wave--mean-flow interactions, which Plumb \cite{plumb_interaction_1977} used to construct an idealized model.
The model considers the interaction of two counter-propagating gravity waves with the same frequency, wavelength, and amplitude, with a mean-flow. 
This model was realised experimentally using an oscillating membrane at the boundary of a linearly-stratified layer
\cite{plumb_instability_1978,otobe_visualization_1998,semin_generation_2016,semin_nonlinear_2018}.
The experiments can drive oscillating mean-flows similar to the QBO, as predicted by the Lindzen and Holton theory.
More recently, \cite{renaud2019periodicity} simulated the Plumb model numerically to explain the 2016 disruption of the QBO \cite{osprey_unexpected_2016,newman_anomalous_2016}.
Although they find regular oscillations in the mean-flow at low forcing amplitudes, the mean-flow becomes quasi-periodic, and eventually chaotic as the forcing amplitude increases.
Atmospheric forcing amplitudes are in the chaotic mean-flow regime, suggesting that the Plumb model must be refined to explain the QBO.

Because of its influence on weather events, it is crucial that the period and amplitude of the QBO are accurately modelled in Global Circulation Models (GCMs).
Due to their relatively coarse resolution, GCMs cannot compute small time- and length-scale motions like IGWs. Therefore, IGWs are parameterised in order to generate a realistic QBO.
Some GCMs are able to self-consistently generate the QBO \cite{lott_stochastic_2012,lott_stochastic_2013}, which is considered a key test of a model's wave parameterization.
The dependence of the QBO on vertical resolution and wave spectrum properties is not yet understood \cite{xue_parameterization_2012,anstey_simulating_2016,yu_sensitivity_2017}.

Direct Numerical Simulations (DNS) have found mean-flow oscillations generated by a broad spectrum of IGWs self-consistently excited by turbulence \cite{couston_order_2018}.
Because 2D simulations are expensive to run for long integration times, the influence of the forcing on the oscillations could not be studied extensively.
Only a Plumb-like one-dimensional model can realistically allow for a systematic exploration. 

Despite the existence of a broad spectrum of waves in nature, only \cite{saravanan_multiwave_1990} has studied multi-wave forcing. Investigating three different forcing spectra, he found the QBO period is affected by the choice of the spectrum.
In this letter, we consider a wide class of wave spectra in the Plumb model, hence complementing the study of \cite{couston_order_2018}.
We find that forcing a broad frequency range produces regular mean-flow oscillations, even when the forcing amplitude is so large that monochromatic forcing produces a chaotic mean-flow.
This suggests multi-modal forcing is an essential to understand wave--mean-flow interactions.

\section*{Model}
Mean-flow evolution is determined by the spatially-averaged Navier-Stokes equations \cite{bretherton_mean_1969}.
We define the horizontal mean-flow $\overline{u}$; overbar indicates horizontal ($x$) average.
Gravity points in the $-z$ direction, and the velocity fluctuations are $\left(u',w'\right)$.
The horizontal average evolves according to the 1D equation:
\begin{equation}\label{eq:meanflow}
    \partial_t \overline{u} - \nu \partial_{zz} \overline{u} = - \partial_z \overline{u'w'}
\end{equation}
where $\nu$ is the kinematic viscosity.
The mean-flow is forced by the Reynolds stress term on the right-hand side of (\ref{eq:meanflow}).
In the Plumb model, the Reynolds stress comes from the self-interaction of IGWs.
We excite the waves at the bottom boundary $z=0$, and propagate the waves through a linearly-stratified domain characterized by a fixed buoyancy frequency $N$.
We non-dimensionalize the problem by setting the top boundary at $z=1$ and setting $N=1$.
Wave damping leads to vertical variation in the Reynolds stress, driving the mean-flow.

We consider a superposition of waves $\psi_i=A_i(z)e^{i\left(k_xx-\omega_i t\right)}$, where $\psi_i$ is the streamfunction, $k_x$ the horizontal wavenumber, and $\omega_i$ the angular frequency.
Assuming a time-scale and length-scale separation between the fast, short scale IGWs, and the slowly evolving, long scale mean-flow, we use the WKB approximation to derive an expression for $A_i(z)$.
We also make use of the following approximations. We take the ``weak'' dissipation approximation, assume the background stratification is constant in space and time, and neglect wave-wave nonlinearities, except when they affect the mean-flow; see details in \cite{renaud2020holton} and \cite{suppmat}. Unlike the classical model which uses the hydrostatic approximation \cite{plumb_interaction_1977, renaud2020holton}, we solve the full vertical momentum equation for the wave, which allows for high-frequency IGWs. We  neglect Newtonian cooling but consider  diffusion of the stratifying agent (with diffusivity $D$), which is relevant for both laboratory experiments and the DNS described below. The inverse damping lengthscale is given by 

\begin{equation}
l_i^{-1}=\frac{1}{2k_x}\times \frac{\nu+D}{\left(1-k_x^2c_i^2\right)^{1/2}c_i^4}
\end{equation}

where $c_i = \frac{ \omega_i \mp k_x \overline{u}}{k_x}$; $\mp$ accounts for the wave direction of propagation.
The right-hand side of (\ref{eq:meanflow}) is written as a sum of independent forcing terms $\sum_i \tilde{F_i}(z)$, where $\tilde{F_i}(z)$ is related to the amplitude $A_i(z)$ for each wave.

\begin{figure*}[t]
    \centering
    \includegraphics[scale=.6]{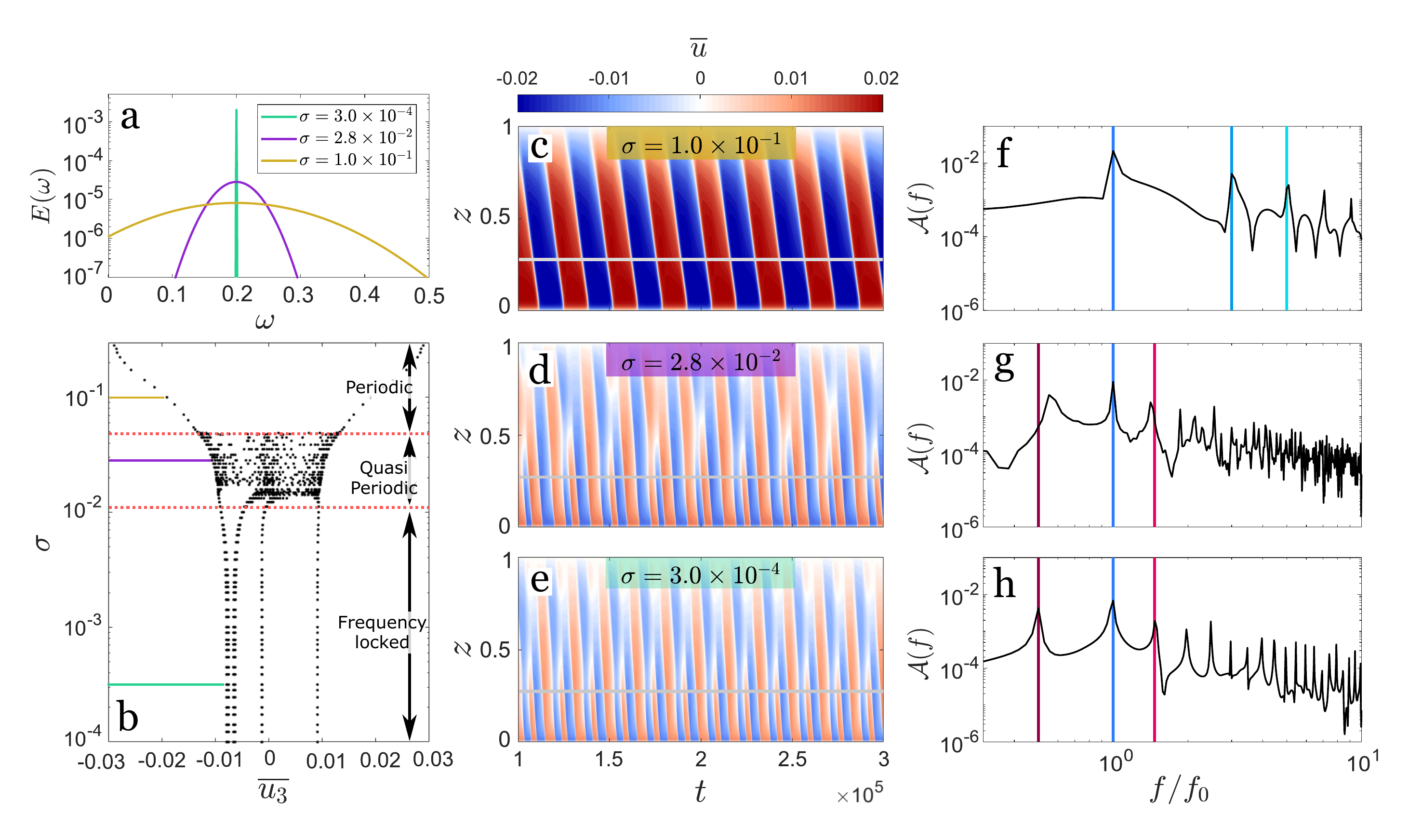}
    \caption{Mean-flows generated by a Gaussian forcing spectrum with standard deviation $\sigma$, central frequency $\omega_0=0.2$ and total kinetic energy $E_{tot} = 2\times10^{-6}$. (a) Example forcing spectra: $\sigma=1\times10^{-1}$ in brown,  $\sigma=2.8\times10^{-2}$ in purple and $\sigma=3\times 10^{-4}$ in green. (b) Poincar\'e map: each dot is the value of the flow $\overline{u}\left(z,t\right)$ at $z=0.5$ (denoted $\overline{u_3}$) when $\overline{u}\left(z=0.1\right) = 0$. Periodic oscillations are represented by two points, symmetric about $\overline{u_3} = 0$ (see \cite{suppmat}). The asymmetry for the points below $\sigma \sim 10^{-2}$ is due to the initial condition. (c-e) Hovm\"oller diagrams of the mean-flow $\overline{u}$ for different $\sigma$. (f-h) Amplitudes of the temporal Fourier transform taken at $z=0.25$ for each Hovm\"oller diagram. Frequencies are normalised by the frequency of maximum amplitude $f_0$, indicated by the dark blue line. The dark red line shows $f=f_0/2$, and the other vertical lines represent harmonics.}
    \label{fig:variance}
\end{figure*}

The classic Plumb model considers a single value of $\omega$, $k_x$, and forcing amplitude $A(z=0)$.
To account for the multi-modal excitation of waves in natural systems, we consider excitation by multiple frequencies with different forcing amplitudes.
The kinetic energy of the waves is given by $E_{tot} = \int (dE/d\omega) d\omega$, and we force the waves so the energy density $dE/d\omega$ is a Gaussian in frequency centered at $\omega=\omega_0$ with a standard deviation $\sigma$.
We discretize this spectrum with $N_\omega$ standing waves of frequency $\{\omega_i\}$ with frequency-spacing $\omega_{i+1}-\omega_i=\Delta \omega$. The forcing amplitude is $A_i=\sqrt{2(dE/d\omega)\Delta\omega \frac{\omega_i^2}{k_x^2}}$.
Our results depend only on $dE/d\omega$, not on the amplitudes of the individual modes (which vary with $N_\omega$).
We only consider a single value of $k_x = 4\pi$.
This general forcing spectrum allows us to study the transition from monochromatic forcing ($\sigma\rightarrow 0$) to multi-modal forcing (white noise in the limit $\sigma\rightarrow\infty$) so we can compare our results with past monochromatic studies. We consider the dimensionless dissipation $\nu =  2.8\times 10^{-6}$ and $D=\nu/700$, estimated from laboratory experiments \cite{semin_nonlinear_2018}; we  explored the ranges $\omega_0 \in [0.1;0.55]$, $\sigma \in [10^{-4}; 3\times 10^{-1}]$ and $E_{tot} \in [7\times 10^{-7}; 10^{-5}]$. For comparison with previous monochromatic studies \cite{plumb_interaction_1977, renaud2019periodicity}, the wave forcing Reynolds number of each individual wave goes up to $\sim 100$, which is comparable to the range explored in \cite{renaud2019periodicity}. Additional simulations were also performed with spectra representative of turbulence. Some of the results are discussed at the end of this letter.
We initialize $\overline{u}$ with a small amplitude sinusoid.
Boundary conditions for the mean-flow are no-slip ($\overline{u}=0$) at the bottom and free-slip ($\partial_z\overline{u}=0$) at the top.
The waves freely propagate out of the domain's top boundary without reflection.
Section A.2 of \cite{suppmat} describes our spatial and temporal discretization, and demonstrates numerical convergence.

We investigated the influence of top boundary conditions (BCs) on the mean-flow evolution using two-dimensional DNS of the Navier-Stokes equations \cite{dedalus_burns_2016,dedalus_burns_2019}. We found the top BCs only marginally influence the period and amplitude of the oscillations, and do not affect their dynamical regime \cite{suppmat}.
We thus focus on results from our 1D model, which allows for the systematic exploration of a larger parameter space.
The vertical extent of the simulation domain does not qualitatively change our results, even though some high-frequency waves have attenuation lengths greater than the domain height \cite{suppmat}.


\section*{Results}

We investigate the influence of the forcing bandwidth on the mean-flow evolution, varying the standard deviation $\sigma$ of our Gaussian excitation spectrum, but fixing the central frequency and total energy. Figure \ref{fig:variance} shows that for narrow distributions $10^{-4} \leqslant \sigma <  10^{-2}$, the system produces frequency-locked oscillations, with slow oscillations in the upper part of the domain and fast oscillations in the bottom part (Figure \ref{fig:variance}e).
The frequency power spectrum for  $\sigma = 3 \times 10^{-4}$ (Figure \ref{fig:variance}h) shows peaks at these two frequencies ($f_0$ and $f_0/2$), as well as at harmonic/beating frequencies.

At $\sigma \approx 10^{-2}$, the oscillations transition from a frequency-locked regime to a quasi-periodic regime (Figure \ref{fig:variance}d,\,g).
A second bifurcation to periodic oscillations occurs at $\sigma = 5 \times 10^{-2}$, with only one dominant frequency (plus harmonics) appearing in the corresponding spectrum (Figure \ref{fig:variance}f).
Forcing spectra with wide bandwidths lead to more organised, QBO-like states.

A wide bandwidth forcing spectrum includes more frequencies; naively, this would lead to chaotic mean-flows.
However, a wider spectrum also excites higher frequency waves.
High-frequency waves experience less damping than low-frequency waves, so they can propagate higher.
Because the QBO reversal occurs at the top, we hypothesize the period and regularity of the oscillation is determined by the highest frequency wave above a threshold amplitude.
At fixed wave forcing amplitude, higher frequency waves correspond to more regular oscillations with longer periods (Figure 2a).
Furthermore, the amplitude of the mean-flow is larger when forced by higher frequency waves, because the phase velocity is larger.
This means the amplitude of the mean-flow is larger for the periodic oscillations forced by a wide spectrum (Figure 1b).

\begin{figure}[t]
    \centering
    \includegraphics[scale=.45]{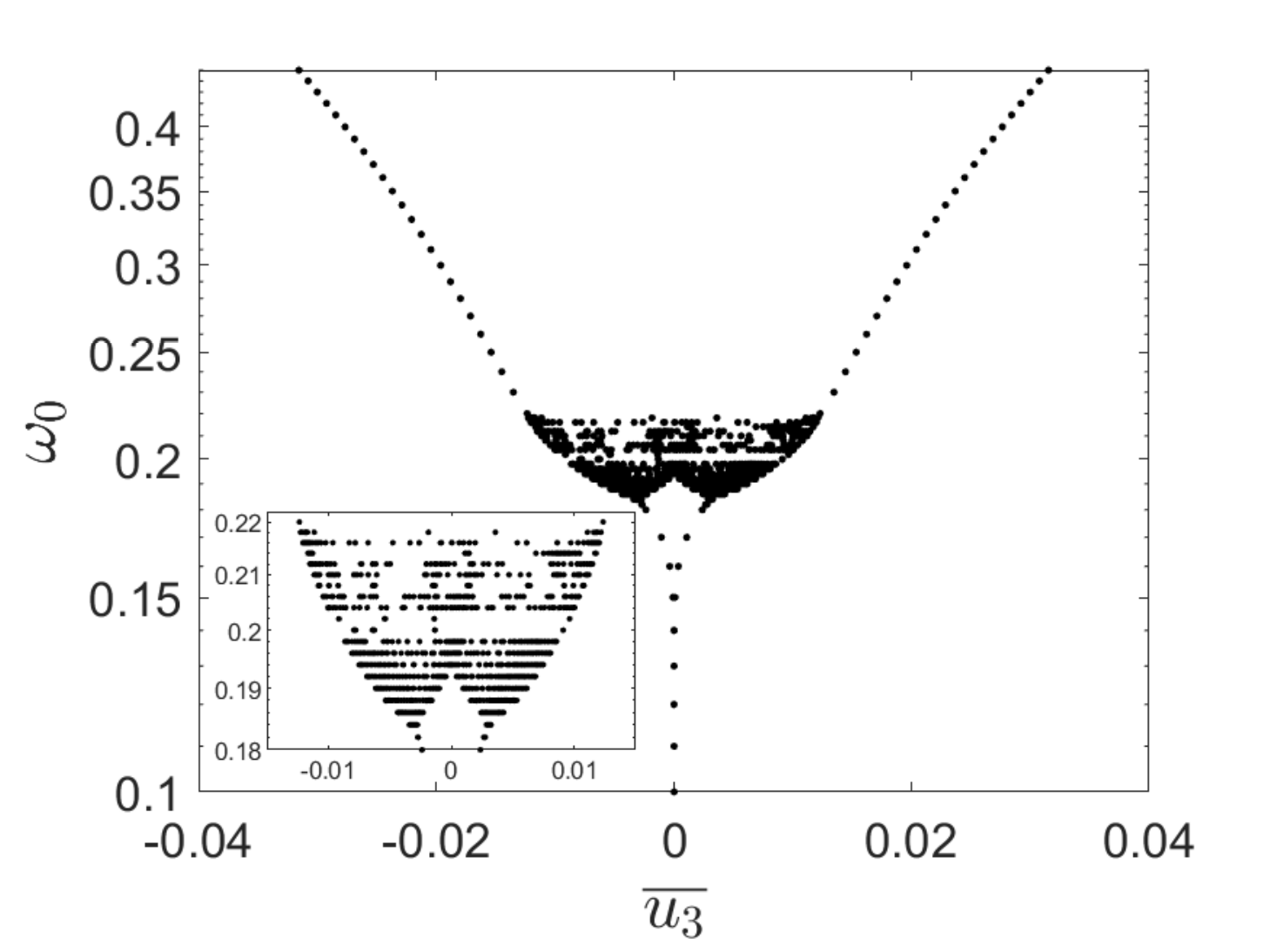}
    \hspace{.1pt}
    \includegraphics[scale=.45]{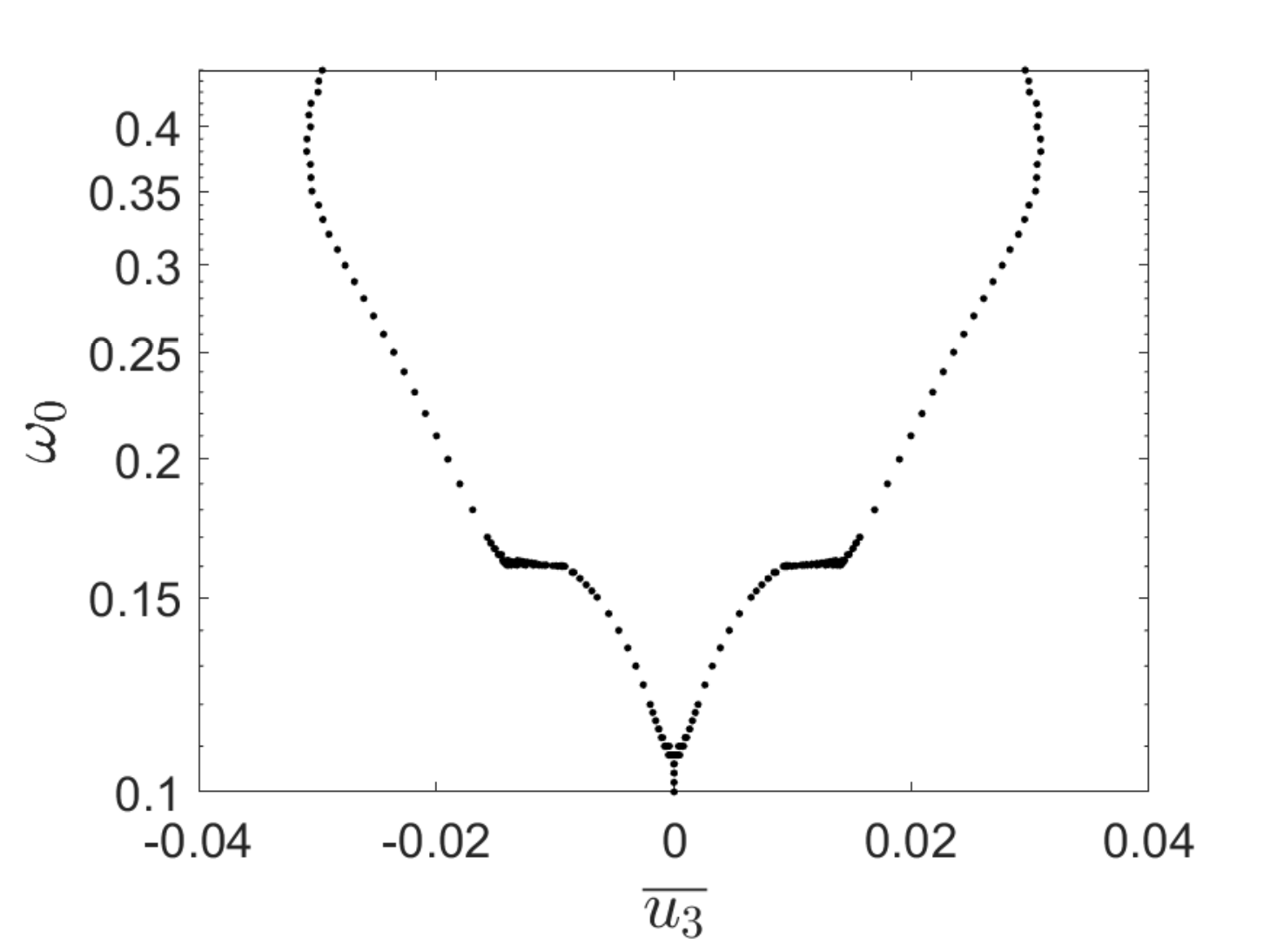}
    \caption{Poincar\'e maps using the same method as Figure \ref{fig:variance}b for simulations with different $\omega_0$, with $E_{tot} = 2\times10^{-6}$ and $\sigma = 10^{-4}$ (top) or $\sigma = 10^{-1}$ (bottom). The inset zooms in on the frequency range $\omega_0 \in [0.18; 0.22]$ to show the transitions between periodic, frequency-locked, and quasi-periodic regimes.}
    \label{fig:mu_var}
\end{figure}
Figure \ref{fig:mu_var} shows Poincar\'{e} maps for simulations forced with different central frequencies $\omega_0$, at fixed total energy $E_{tot}$.
We used either a narrow distribution ($\sigma=10^{-4}$) or a broad distribution ($\sigma=10^{-1}$).
The top panel  ($\sigma=10^{-4}$, similar to monochromatic forcing) shows a transition from periodic oscillations to non-periodic oscillations at $\omega_0 = 0.184$.
There is a second bifurcation at $\omega_0 = 0.22$ leading to periodic oscillations.
The amplitude of the oscillations rises with $\omega_0$ because the phase velocity increases.
The Poincar\'{e} map for variable $\omega_0$ (Figure 2) is qualitatively similar to the Poincar\'{e} map for variable $\sigma$ (Figure 1b), suggesting that the primary effect of increasing $\sigma$ is to put more power into high-frequency waves.
Note, the frequency of the second bifurcation to periodic oscillations changes with domain height because the forced wave has a viscous attenuation length greater than our domain height. However, the periodic oscillations for large $\sigma$ are not due to the finite domain size \cite{suppmat}.
The bottom panel of Figure 2 ($\sigma=10^{-1}$, wide spectrum) shows periodic oscillations for all $\omega_0$.
Once again, we find that a wide forcing spectrum with many frequencies will almost always generate regular, periodic oscillations.

\begin{figure}[t]
    \centering
    \includegraphics[scale=.45]{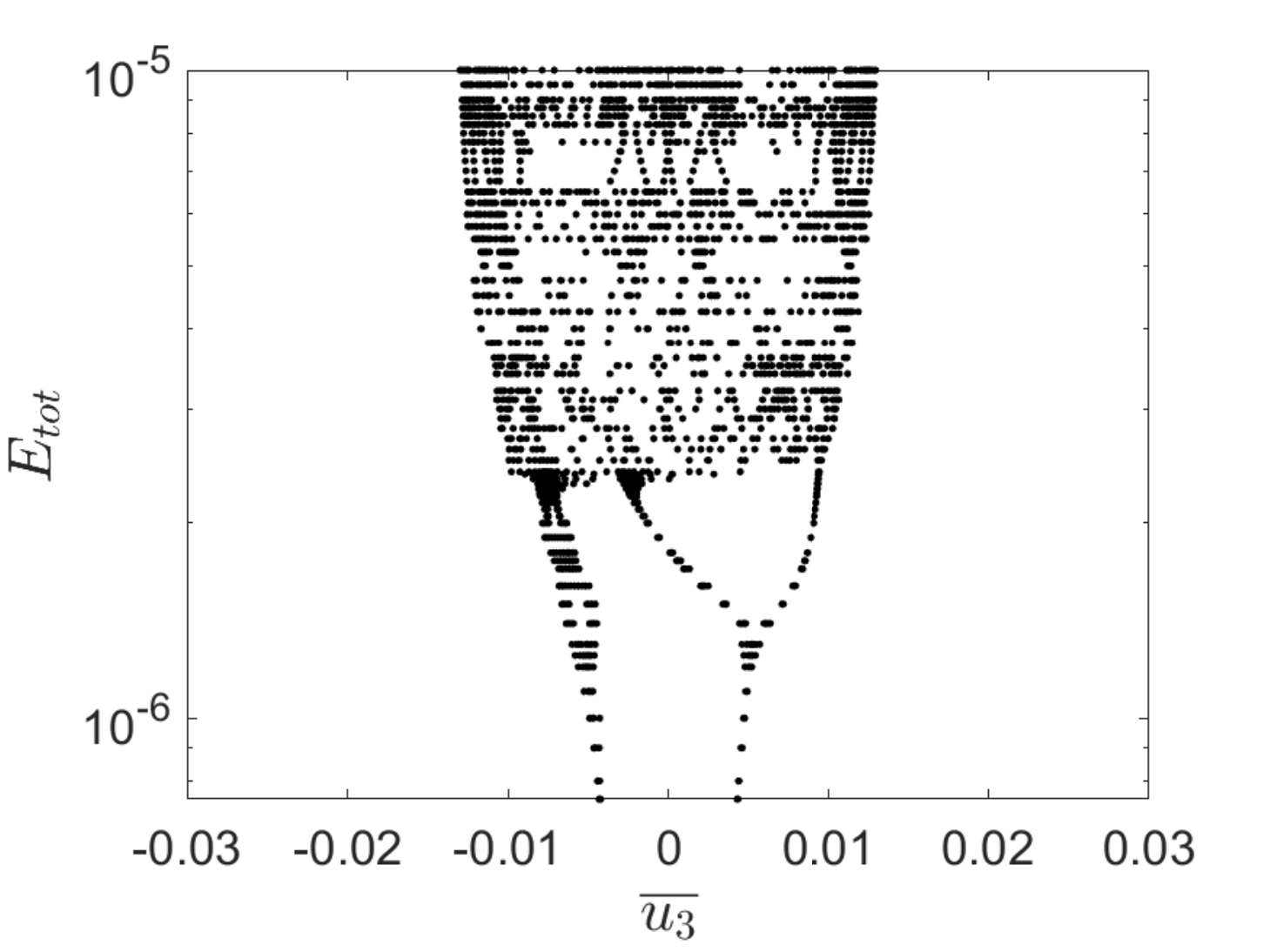}
    \hspace{.1pt}
    \includegraphics[scale=.45]{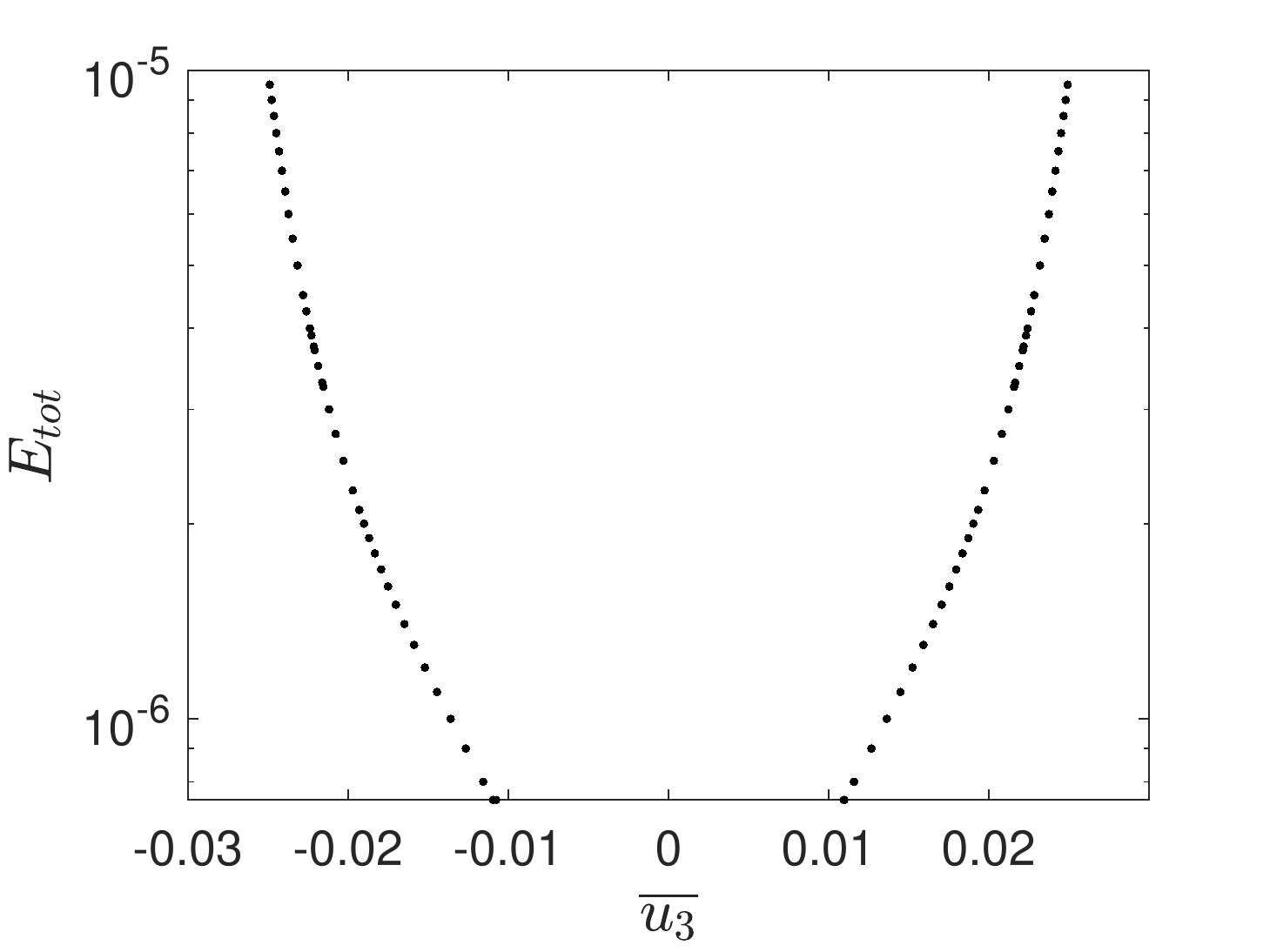}
    \caption{Poincar\'e maps using the same method as Figure \ref{fig:variance}b for simulations with different $E_{tot}$, with $\omega_0=0.2$ and $\sigma = 10^{-4}$ (top) or $\sigma = 10^{-1}$ (bottom).}
    \label{fig:energy}
\end{figure}

In Figure 3, we plot Poincar\'{e} maps for different forcing amplitudes, with fixed $\omega_0$ and $\sigma$.
The top panel ($\sigma=10^{-4}$, almost monochromatic forcing) qualitatively reproduces the results from \cite{renaud2019periodicity}.
As the amplitude of the monochromatic forcing increases, periodic oscillations bifurcate into frequency-locked oscillations ($E_{tot}\approx 1.4\times 10^{-6}$), and then again into quasi-periodic or chaotic oscillations ($E_{tot}\approx 2.4\times 10^{-6}$).
On the other hand, when forcing with a wide spectrum ($\sigma=10^{-1}$, bottom panel), we only find regular periodic oscillations.


\section*{Discussion}
Our study demonstrates that a broad spectrum of IGWs can generate regular mean-flow oscillations.
Whereas large-amplitude monochromatic forcing often generates chaotic mean-flows \cite{renaud2019periodicity}, forcing a broad spectrum of waves consistently generates periodic mean-flows, similar to what is observed in the Earth's atmosphere.
The mean-flow evolution appears to be determined by high-frequency waves which can propagate higher, and control the mean-flow's reversal.
We hypothesize the disruption observed in 2016 \cite{osprey_unexpected_2016,newman_anomalous_2016} is due to intense events which focused significant energy into waves with similar frequency and wavenumber.
Those waves could then trigger non-periodic reversals for a short time \cite{renaud2019periodicity}.

We have run several simulations with forcing spectra more representative of turbulence.
In these simulations, we assume $dE/d\omega$ is constant for $\omega<\omega_c$, and decreases as a power-law for $\omega>\omega_c$.
We tested an $\omega^{-5/3}$ power-law corresponding to Kolmogorov's law and an $\omega^{-3}$ power-law corresponding to the energy cascade observed in rotating turbulence.

\begin{figure}
    \centering
    \includegraphics[scale = .5]{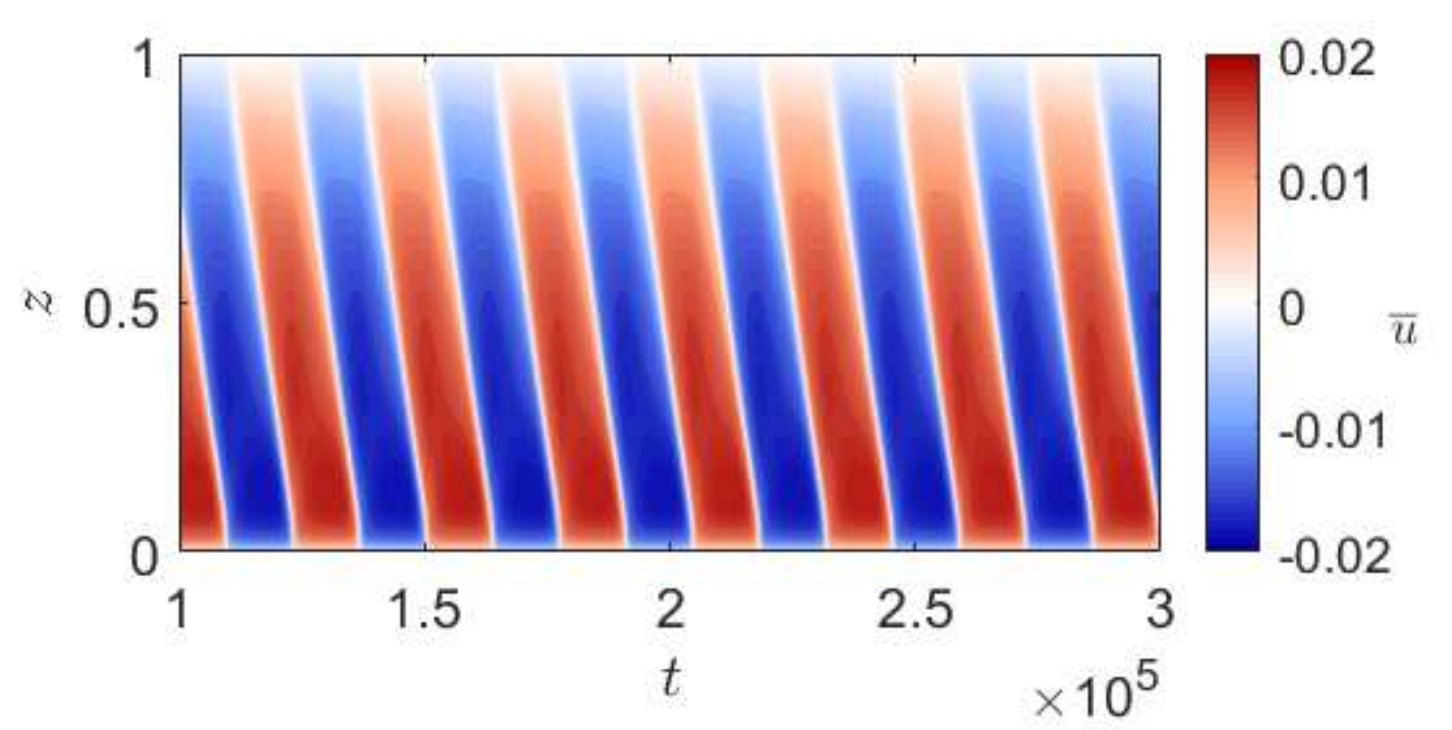}
    \includegraphics[scale = .5]{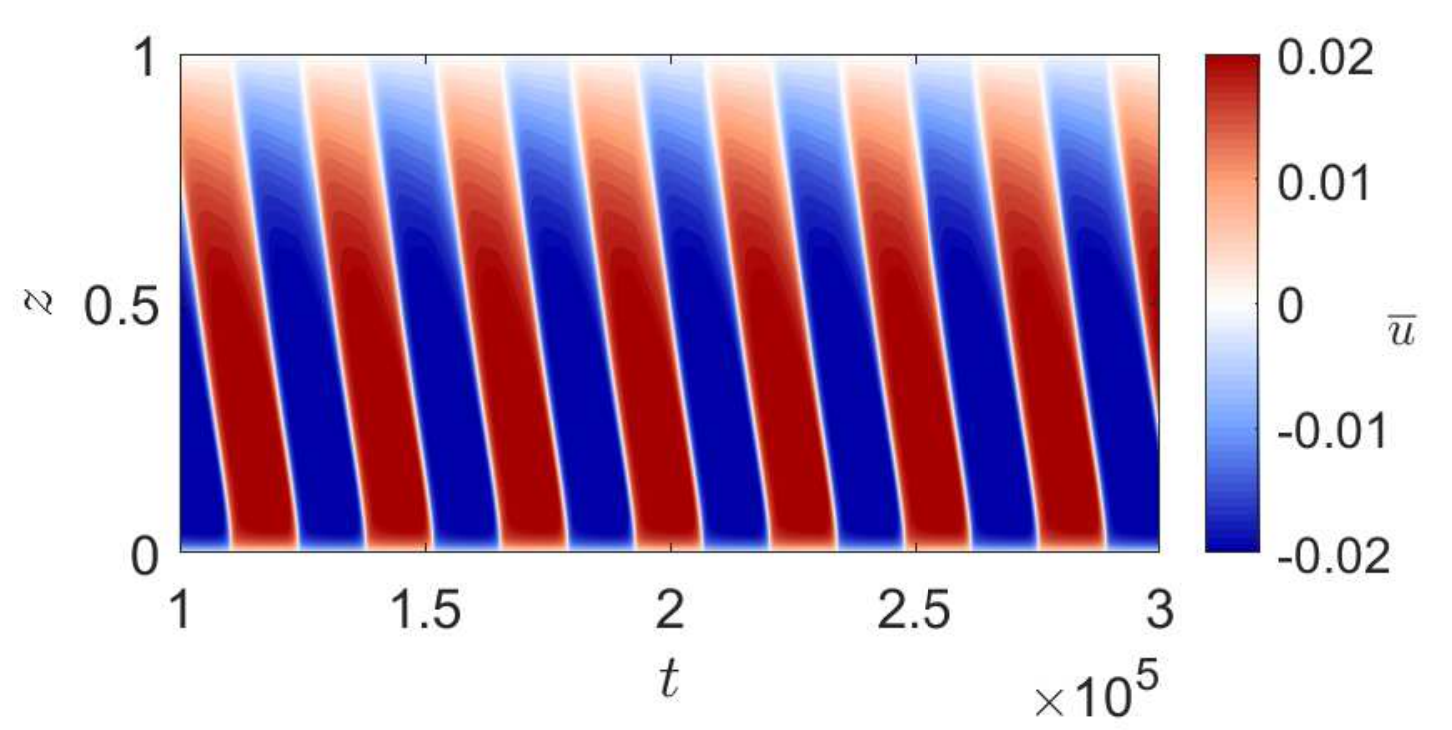}
    \caption{Hovm\"oller diagram of the mean-flow $\overline{u}$ in simulations with different forms of the forcing spectrum.
    In the top panel, $dE/d\omega$ is constant up to $\omega_c=0.2$, and decreases as $\omega^{-5/3}$ for higher frequencies; the total energy is $E_{tot}=2.5\times10^{-6}$. In the bottom panel, $dE/d\omega$ is constant up to $\omega_c=0.23$, and decreases as $\omega^{-3}$ for higher frequencies; the total energy is $E_{tot}=2.8\times10^{-6}$.}
    \label{fig:forcage_spectre}
\end{figure}

We find that different forcing spectra can lead to the same mean-flow evolution.
The top panel of Figure \ref{fig:forcage_spectre} shows a Hovm\"oller diagram for the $\omega^{-5/3}$ spectrum with $E_{tot} = 2.5\times10^{-6}$.
It is quantitatively similar to the Hovm\"oller diagram of Figure \ref{fig:variance}c, obtained with a Gaussian forcing and $20\%$ less energy (oscillation periods differ by $4.5\%$ and amplitudes by $15\%$). 
The $\omega^{-3}$ spectrum with $E_{tot} = 2.8\times10^{-6}$ also generates a similar mean-flow (bottom panel of Figure \ref{fig:forcage_spectre}; oscillation periods are equal and amplitudes differ by $14\%$).
Since multiple wave spectra can produce the same mean-flow oscillations, reproducing the Earth's QBO in a GCM does not mean the IGW parameterization is correct.


Our simulations use parameters similar to laboratory experiments of the QBO \cite{plumb_instability_1978, semin_nonlinear_2018}.
In the atmosphere,  wave attenuation also occurs via Newtonian cooling, which we did not include in our simulations. Besides, the forcing is stronger and viscosity is weaker.
Waves deposit their energy either via critical layers (which we include in our model), or via breaking due to wave amplification from density variations.
Our calculations show the mean-flow period and amplitude is set by high-frequency waves with large viscous attenuation lengths.
Although many more atmospheric waves have small viscous attenuation lengths, high-frequency waves are still more important than low-frequency waves because they do not encounter critical layers.
Thus, we believe high-frequency waves likely play a key role in setting the QBO properties, just as they are important in our simulations. 


In conclusion, this letter shows that the frequency spectrum of internal gravity waves plays a key role in the generation and properties of periodic large-scale flow reversals like the QBO.
Although we studied general frequency spectra, we limited our investigation to a single wavenumber.
Future work should also include the wide range of horizontal wavenumbers ($10-1000$~km) observed in the atmosphere, including low-frequency, planetary-scale waves which may also be an important source of momentum for the QBO \cite{dunkerton_role_1997}.
Additionally, the study of the QBO in a fully coupled, convective--stably-stratified model system by \cite{couston_order_2018} showed that accounting for only the energy spectrum in a Plumb-like model is not sufficient to reproduce the realistic reversals; it also requires information about higher order statistics.
Clearly, reliable parameterization of this climatic metronome in GCMs still demands additional work.

\section*{Acknowledgment}
The authors acknowledge funding by the European Research Council under the European Union's Horizon 2020 research and innovation program through Grant No. 681835-FLUDYCO-ERC-2015-CoG. They also thank Benjamin Favier (IRPHE, CNRS, Marseille, France) for fruitful discussions. DL is funded by a Lyman Spitzer Jr.~fellowship.



\pagebreak
\newpage
\setcounter{equation}{0}
\setcounter{figure}{0}
\setcounter{table}{0}
\setcounter{page}{1}
\makeatletter
\renewcommand{\theequation}{S\arabic{equation}}
\renewcommand{\thefigure}{S\arabic{figure}}
\renewcommand{\bibnumfmt}[1]{[S#1]}
\renewcommand{\citenumfont}[1]{S#1}

\begin{widetext}
\begin{center}
    \textbf{\large Supplemental Material for Multi-modal excitation to model the Quasi-Biennial Oscillation}\\[15pt]
    Pierre L\'eard$^1$, Daniel Lecoanet$^2$, Michael Le Bars$^{1}$\\[3pt]
    $^1$ \textit{ Aix Marseille Universit\'e, CNRS, Centrale Marseille, IRPHE, Marseille, France\\
$^2$ Northwestern University, Engineering Sciences and Applied Mathematics, Evanston, IL, 60208}\\
    (Dated: \today )
\end{center}

\subsection{Model Details}
\subsubsection{Equations}
The generation of a mean-flow by internal gravity waves (IGWs) can be numerically investigated with the two-dimensional Navier-Stokes equations, horizontally averaged along the $x$ direction. The horizontally-averaged momentum equation gives the mean-flow equation along the $z$ axis,
\begin{equation}
    \partial_t \overline{u} - \nu \nabla^2 \overline{u} = -\partial_z \overline{u'w'} \label{eq:meanflow_s}
\end{equation}
where $\overline{u}$ is the horizontally averaged horizontal velocity, and $\left(u',w'\right)$ are the horizontal and vertical velocity fluctuations. The right-hand side of equation (\ref{eq:meanflow_s}) is the forcing term and is computed from the linearized wave equation.
We use the quasi-linear approximation, which neglects non-linear effects except those representing wave--mean-flow interactions \cite{plumb_momentum_1975_s}. After non-dimensionalization by the vertical extent of the domain $H$ and the buoyancy frequency $N$, which is assumed to be constant in time, one can write the wave equation for the streamfunction $\psi$ as
\begin{equation}\label{eq:waveeq}
        \left( \partial_t + \overline{u} \partial_x - D \nabla^2  \right) \left[ \left( \partial_t + \overline{u} \partial_x - \nu \nabla^2 \right) \nabla^2 - \overline{u}_{zz} \partial_x \right] \psi = \partial_x^2 \psi
\end{equation}
where $D = \frac{D^*}{H^2N}$ is the dimensionless diffusivity (thermal or molecular, depending on the physical origin of the stratification) and $\nu = \frac{\nu^*}{H^2N}$ is the dimensionless kinematic viscosity.
Using the Wentzel-Kramers-Brillouin (WKB) approximation, the streamfunction $\psi$ is
\begin{equation}\label{eq:ansatz}
    \psi_\pm(z) = A_0 e^{\int_z i \varphi_\pm(z') + \chi_\pm(z') dz'}e^{i\left(k_x x \mp \omega_\pm t\right)} + cc
\end{equation}
where $k_x$ is the horizontal wavenumber, $\omega_\pm =  \omega \mp k_x \overline{u}$ is the Doppler-shifted frequency, and $\varphi$ and $\chi$ are two functions of $z$.
We assume we are in the weak-dissipation limit (we neglect the $D\nu\nabla^6$ term) and that there is a time- and length-scales separation between the waves and the mean-flow ($\partial_t \overline{u} \sim 0$ and $\overline{u}_{zz} \sim 0$).
$\varphi$, $\chi$, and the Doppler-shifted frequency $\omega_\pm$ are all slowly varying functions of $z$, due to the $z$ dependence of $\overline{u}$.

Using the ansatz (\ref{eq:ansatz}) and the wave equation (\ref{eq:waveeq}), together with the approximations described above, we can derive equations for $\varphi$ and $\chi$,
\begin{align}
\varphi_\pm^2 = k_x^2 \frac{1-\omega_\pm^2}{\omega_\pm^2}, && \chi_\pm = - \frac{1}{2} \frac{\mathrm{d}}{\mathrm{d}z} \mathrm{ln} \varphi_\pm -\frac{\left(\nu + D\right)k_x^3}{2\left(1-\omega_\pm^2\right)^{1/2}\omega_\pm^4}.
\end{align}
$\varphi$ is the vertical wavenumber and satisfies the inviscid dispersion relation for IGWs. $\chi$ includes the vertical attenuation of the wave due to the dissipative processes.

Finally, the self-interaction forcing term for each wave is
\begin{equation}\label{eq:forcingterm}
    -\partial_z \overline{u'w'}^\pm = \pm \frac{2A_0^2 \varphi_0(z=0)\left(\nu + D\right)k_x^4 }{\left(1-\omega_\pm^2\right)^{1/2}\omega_\pm^4} \times e^{-\int \mathrm{d}z' \frac{\left(\nu + D \right)k_x^3}{\left(1-\omega_\pm^2\right)^{1/2}\omega_\pm^4}}
\end{equation}

To develop the model, several hypotheses are made (weak dissipation, $\overline{u}_{zz}$ is small, time scale separation between the waves and the mean-flow), which may be violated in the course of our simulations.
The weak dissipation limit tends to overestimate dissipation when Doppler-shifted frequencies $\omega_\pm \rightarrow 0$ \cite{lecoanet_numerical_2015_s}.
The forcing term becomes large when $\omega_\pm\rightarrow 0$, and can greatly accelerate the flow.
Therefore, locally, $\overline{u}(z)$ can vary strongly with $z$.
Nevertheless, these approximations are useful for obtaining an analytical expression for the forcing term, and appear to be sufficient to model mean-flow reversals \cite{plumb_instability_1978_s}. 


\subsubsection{Numerical model}
The mean-flow $\overline{u}$ is solved with a semi implicit/explicit time scheme using finite differences.
Time scheme is an order 1 Euler scheme and the laplacian term is solved \textit{via} an order 2 centred finite differences scheme.
At each time step, the forcing term $\partial_z \overline{u'w'}$ is computed from equation (\ref{eq:forcingterm}) at a time $t$. Then, the mean-flow $\overline{u}(z,t+\delta_t)$ is calculated by adding the effect of the divergence of the Reynolds stress and viscosity.
We use a no-slip boundary condition at the bottom, $\overline{u}|_{z=0}=0$, and a free-slip boundary condition at the top $\left. \frac{\partial \overline{u}}{\partial z}\right|_{z=1} = 0$ (see section \ref{sec:top_bc} for a discussion of the top boundary condition).

\begin{figure*}[t]
    \centering
    \includegraphics[scale = .7]{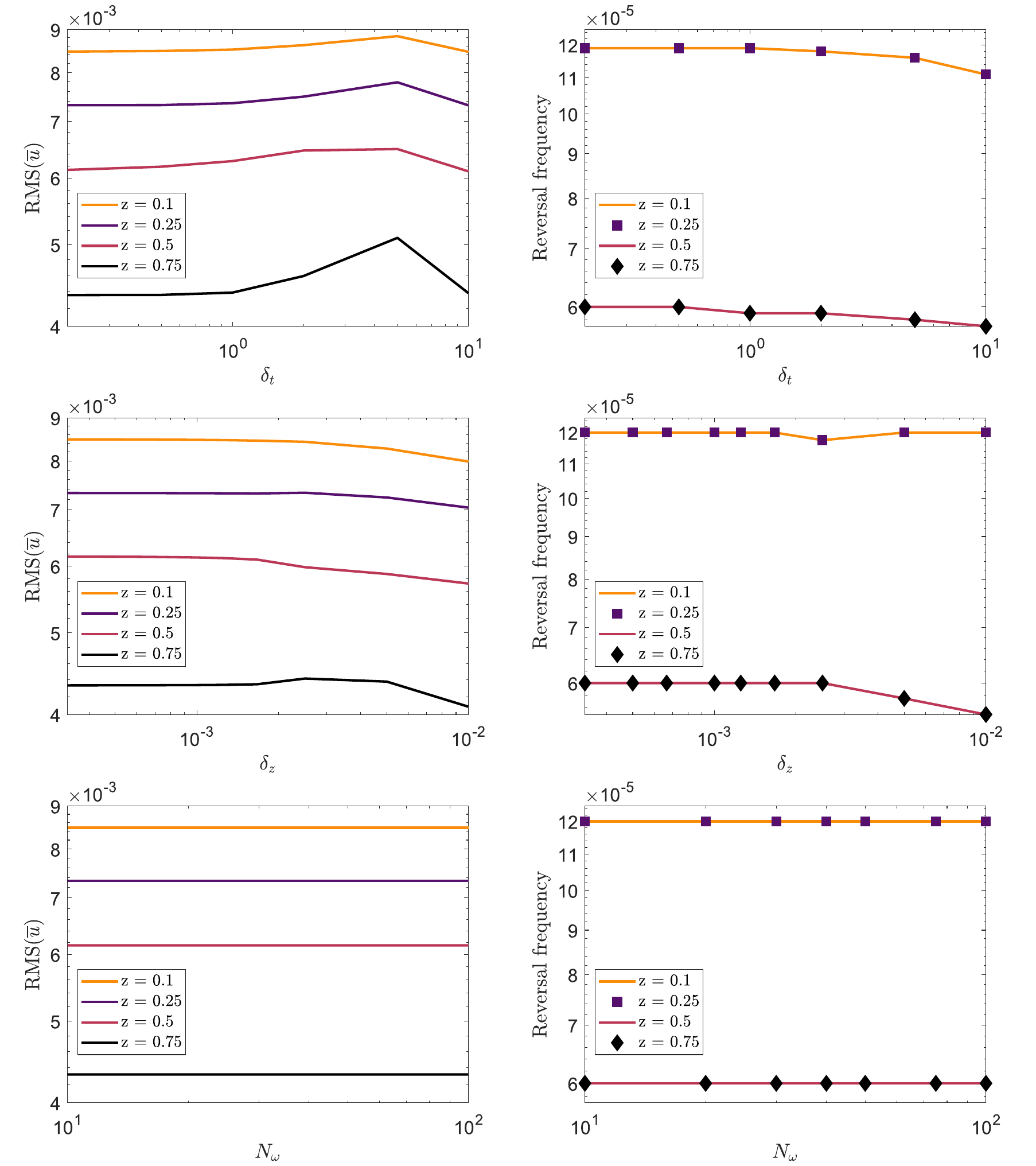}
    \caption{Mean-flow properties for $\omega_0=0.2$, $\sigma = 1\times 10^{-4}$, and $E_{tot} = 2\times 10^{-6}$. Convergence study for $\delta_t$ (top), $\delta_z$ (middle), and $N_\omega$ (bottom).}
    \label{fig:conv_sigma}
\end{figure*}

Singularities appear in equation (\ref{eq:forcingterm}) when  $\omega_\pm \rightarrow 0$ or $\omega_\pm \rightarrow 1$. The former case physically corresponds to a critical layer. Numerically, critical layers are treated as follows: if the mean-flow reaches a value higher than $0.95 \ \omega/k_x$ at a given height $z_c$, the momentum contained in the waves is deposited at the grid point below $z_{c}$ and the forcing terms for $z \geqslant z_c$ are set to $0$. The latter case corresponds to a wave whose Doppler shifted frequency is close to the buoyancy frequency. This is treated as in the critical layer case, with the condition now being $|\omega_{\pm} | \geqslant  \, 0.95 $.

Viscosity and diffusivity are set to experimental values for salty water, as in experiments \cite{semin_nonlinear_2018_s}.
Using typical laboratory parameters $H = 0.41$~m and $N = 2.16 \ \mathrm{s^{-1}}$, the dimensionless dissipation constants are $\nu = 2.8 \times 10^{-6}$ and $D = \nu / 700$. Thus, viscosity is the main dissipative mechanism for both the large-scale flow and the waves.
Note that $D$ only appears in the wave attenuation while $\nu$ appears in both the wave attenuation and the mean-flow dissipation.

There are three main numerical parameters in our simulations: the time step $\delta_t$, the grid spacing $\delta_z$ and the number of frequencies in the spectrum $N_\omega$.
We ran convergence studies to choose the values of these parameters.
We ran simulations with $E_{tot} = 10^{-5}$ or $E_{tot} = 2\times10^{-6}$ and for the two ``extreme'' standard deviations $\sigma = 10^{-4}$ and $\sigma = 3\times10^{-1}$. Simulations are run for $\sim15$ reversal times.

Figure \ref{fig:conv_sigma} shows the mean-flow properties as a function of the three numerical parameters $\delta_t$, $\delta_z$ and $N_\omega$ for the case $\omega_0 = 0.2$, $\sigma=10^{-4}$ and $E_{tot} = 2\times10^{-6}$.
Top panel of figure \ref{fig:conv_sigma} shows convergence as $\delta_t\rightarrow 0$. Comparing results for $\delta_t = 1$ and $\delta_t = 0.05$ shows that frequencies differ by $<1.7\%$ and the flow amplitudes differ by $<1.5\%$. Moreover, dynamical regimes do not depend on $\delta_t$ for the results presented in the paper. 
Other convergence studies were conducted for $\delta_z$ (see middle panel of figure \ref{fig:conv_sigma}) and $N_\omega$ (bottom panel of figure \ref{fig:conv_sigma}) to determine their best values for the systematic study in $\sigma$ (displayed in figure 1 of the main text). Convergence studies with $\sigma= 3 \times 10^{-1}$ are not shown here.
Additional studies were also performed at $E_{tot} = 10^{-5}$ to find the optimal time, space and frequency steps for the energy systematic study (displayed in figure 3 of the main text).

We use $\delta_z = 1/2000$ for all simulations. A total number of frequencies $N_\omega = 20$ was found to be satisfactory for all cases. These frequencies were separated by $\Delta\omega = (\omega_{max}-\omega_{min})/{N_\omega}$. Frequencies with $dE/d\omega < 10^{-4} \times \mathrm{max}(dE/d\omega)$ were not considered in the forcing spectrum. Thus, $\omega_{min}$ (resp. $\omega_{max}$) is the lowest (resp. highest) frequency considered, with at least $0.01\%$ energy of the most energetic wave.
$\delta_t$ was adjusted for each study: we chose $\delta_t = 1$ for the $\sigma$ and $\omega_0$ systematic studies, while $\delta_t$ was set to either $0.1$, $0.5$ or $1$ depending on the value of $E_{tot}$ for the energy systematic study.

\subsection{Poincar\'e map}

\begin{figure*}[htb]
    \centering
    \includegraphics[scale=.45]{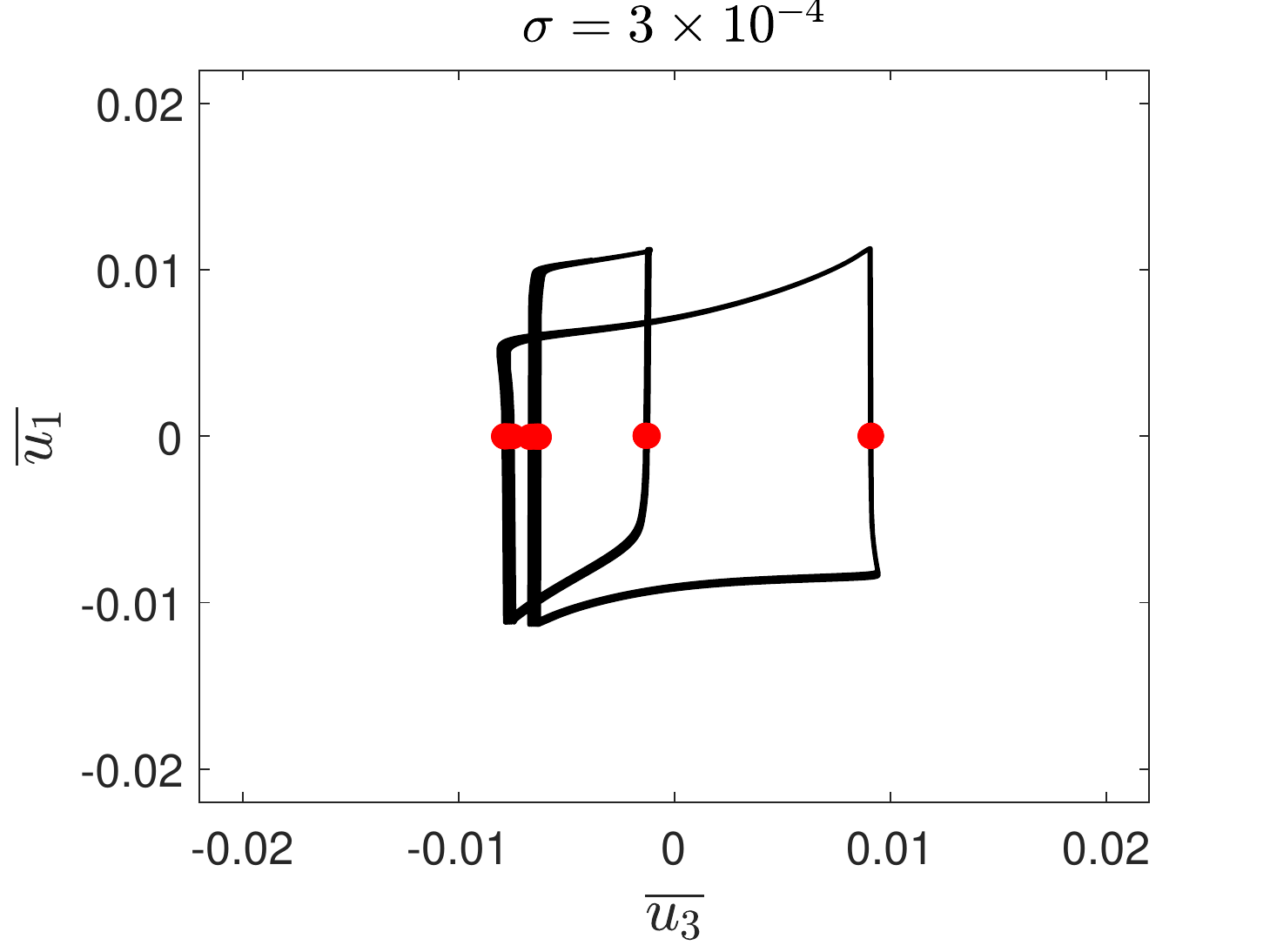}
    \hspace{.1pt}
    \includegraphics[scale=.45]{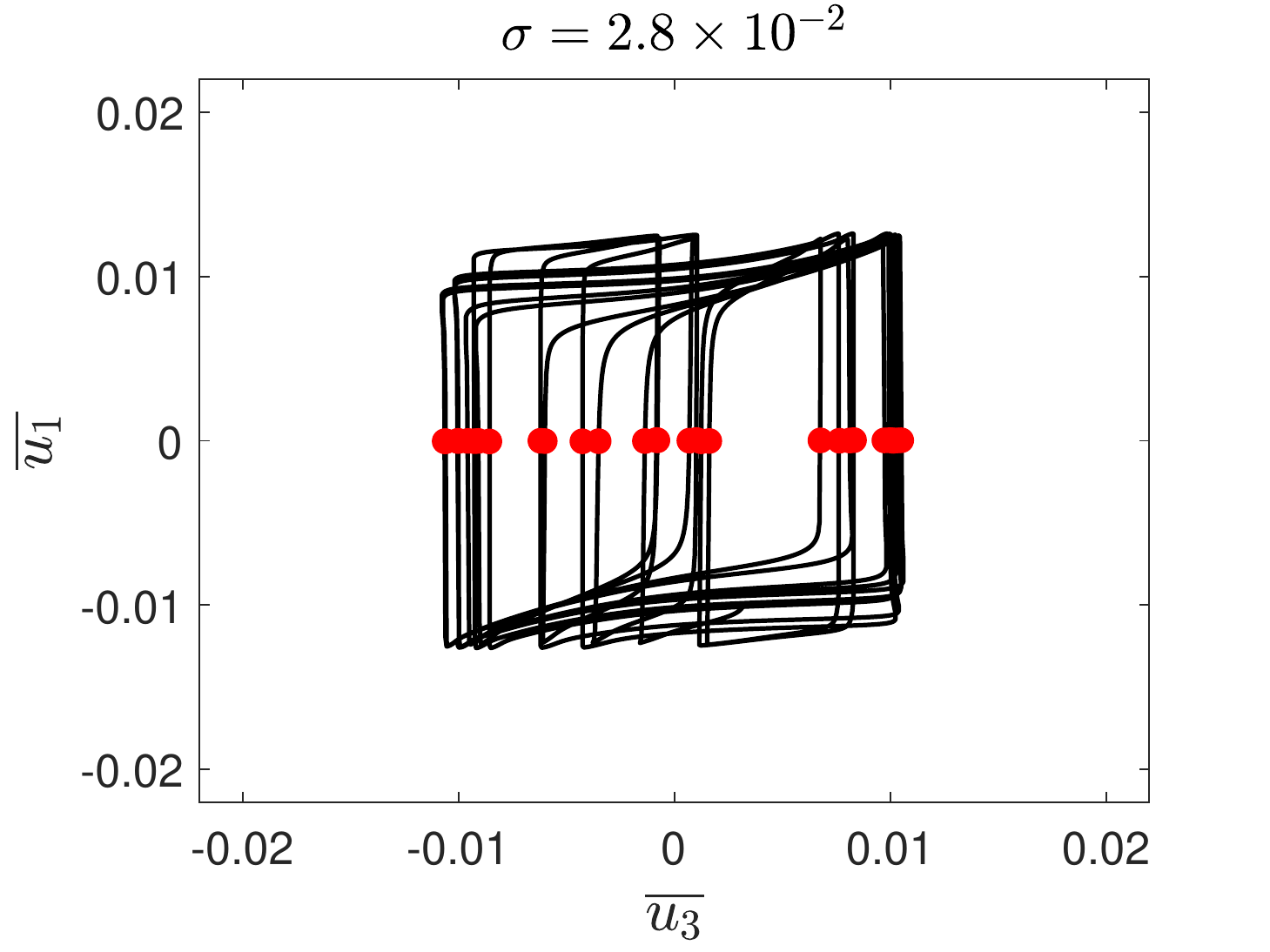}
    \includegraphics[scale=.45]{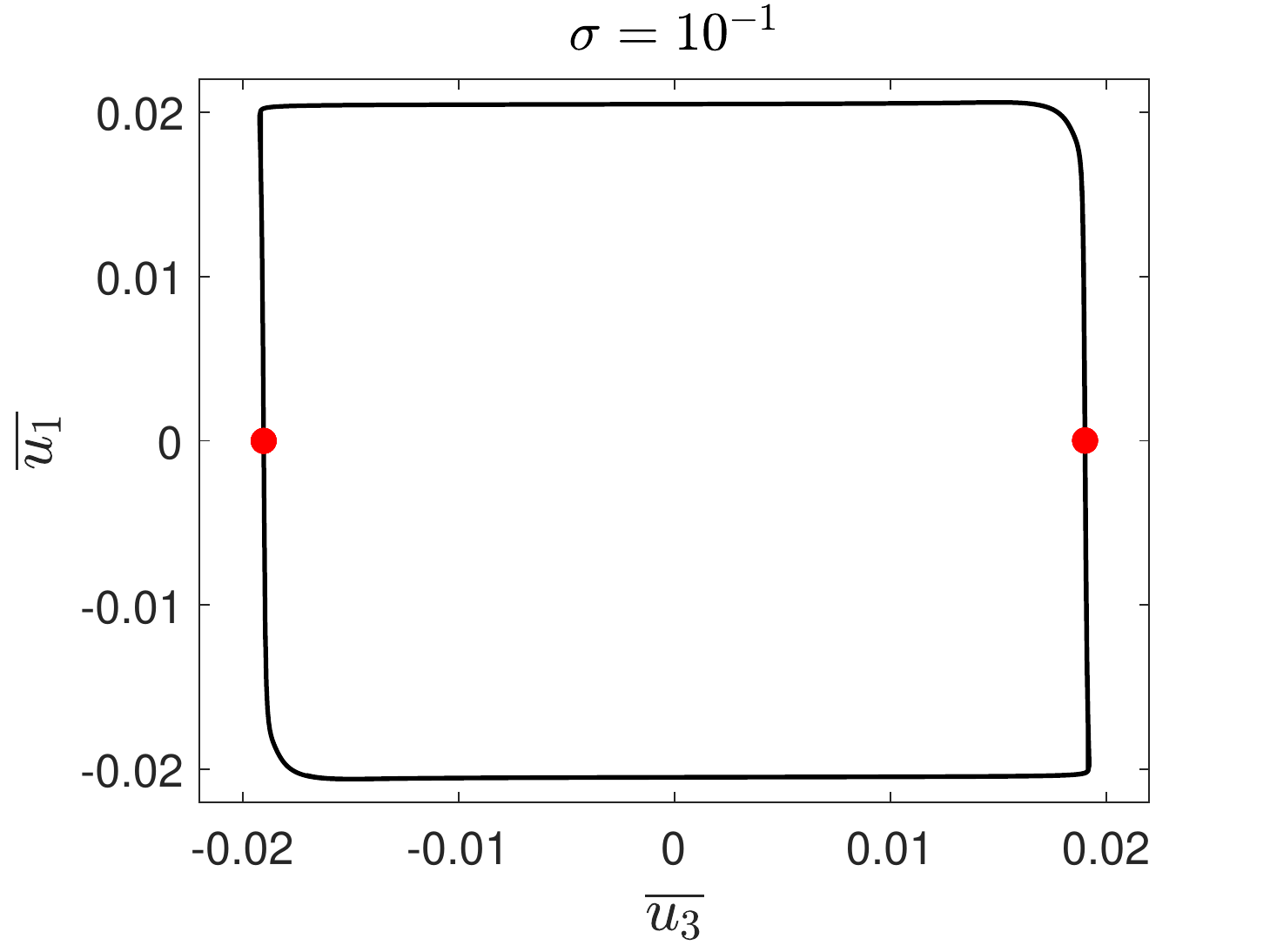}
    \caption{Phase space trajectories on a plane defined by $\overline{u_1} = \overline{u}(z=0.1,t)$ and $\overline{u_3} = \overline{u}(z=0.5,t)$, for three values of $\sigma$ shown in figure 1 of the main text. We fix the central frequency $\omega_0=0.2$ and total energy $E_{tot} = 2\times10^{-6}$. (Top left) Frequency-locked regime for $\sigma = 3\times10^{-4}$. (Top right) Quasiperiodic regime for $\sigma = 2.8\times 10^{-2}$. (Bottom) Periodic regime for $\sigma = 10^{-1}$.}
    \label{fig:u1u3}
\end{figure*}
A Poincar\'e map is a representation of the dynamical behaviour of the system. For each simulation, we first wait for the dynamical system to reach its attractor. We then run for an additional 15 reversal times, and use these data for the Poincar\'e map. Each dot of the map represents the value of the mean-flow at $z=0.5$, denoted $\overline{u_3}$, when the mean-flow at $z=0.1$, denoted $\overline{u_1}$, is equal to $0$. Thus, the set containing the data used to plot the Poincar\'e map is
\begin{equation}
    \mathbb{D} = \{ \overline{u}(z=0.5,t)\ | \ \overline{u}(z=0.1,t) = 0 \}
\end{equation}
Figure \ref{fig:u1u3} shows plots of $(\overline{u_1},\overline{u_3})$. The points in the Poincar\'{e} map are shown with red dots.
Periodic oscillations are characterised by two dots, symmetric about $\overline{u_3}=0$. Frequency-locked oscillations also exhibit a periodic structure with several frequencies appearing in the reversing signal: they are characterised by a finite number of dots.
Finally, for quasi-periodic oscillations, the dots do not superimpose (for a great number of reversals, they would form a continuous line).
The red dots appearing in graphs of figure \ref{fig:u1u3} are the dots appearing in the Poincar\'e map seen in figure 1b of the main text at the corresponding values of $\sigma$.

\subsection{Top boundary conditions}\label{sec:top_bc}

The high-frequency forced waves have viscous attenuation lengths larger than the domain size.
These waves are affected by the vertical size of the domain and the top boundary condition.
In our idealised Plumb-like 1D model, we use a stress-free boundary condition for the mean-flow ($\partial_z\overline{u}=0$), and a radiation boundary condition where waves flow out of the domain.
Here we explore other top boundary conditions.

We compute two-dimensional Direct Numerical Simulations (DNS) using the open spectral solver Dedalus \cite{dedalus_burns_2016_s,dedalus_burns_2019_s}. We solve the following equations:
\begin{subequations}
    \begin{align}
        \partial_t u - \nu \nabla^2 u + \partial_x P &= -(\mathbf{u} \cdot \mathbf{\nabla})u \\
        \partial_t w - \nu \nabla^2 w + \partial_z P - b &= -(\mathbf{u} \cdot \mathbf{\nabla})w \\
        \partial_t b - D \nabla^2 b + w &= -(\mathbf{u} \cdot \mathbf{\nabla})b \label{eq:buoyancy} \\ 
        \mathbf{\nabla \cdot u} &= 0
    \end{align}
\end{subequations}
where $b$ is the buoyancy, $P$ the pressure, and $\nabla = (\partial_x,\partial_z)$. The geometry is two dimensional and periodic along the horizontal axis. We use $n_z = 256$ Chebyshev polynomial functions in the $z$ direction and $n_x = 32$ complex exponential functions in the $x$ direction.
We use $\nu=2.8\times 10^{-6}$ and $D=\nu/7$, representative of thermally stratified water.
To compute nonlinear terms without aliasing errors, we use $3/2$ padding in both $x$ and $z$. For timestepping, we use a second order, two stage implicit-explicit Runge-Kutta type integrator \cite{ascher_implicit_1997_s}, with time steps $\mathcal{O}(10^{-2})$.
Waves with random phases are forced at the bottom domain by setting velocity and buoyancy perturbations at the boundary. We use a forcing spectrum given by $E_{tot} = 2 \times 10^{-6}$, $\omega_0 = 0.2$, $N_\omega = 20$ with either $\sigma = 10^{-1}$ or $\sigma=2.8 \times 10^{-2}$.

We tested four different top boundary conditions:
\begin{itemize}
    \item (i) No-slip: $u|_{z=1} = 0$ and $w|_{z=1} = 0$.
    \item (ii) Free-slip: $\left.\frac{\partial u}{\partial z}\right|_{z=1} = 0$ and $w|_{z=1} = 0$.
    \item (iii) Buoyancy damping layer for $1<z<1.5$ to damp the waves but not the large-scale flow. We implement this by adding a damping term in the right-hand side of equation (\ref{eq:buoyancy}), $-b f(z)/\tau$ where $f(z)=\tanh( (z-1.15)/0.05 )+1)/2$ is a $z$-dependent mask, and the damping timescale $\tau=N^{-1}=1$ in our non-dimensionalization.
    \item (iv) Layer with $N=0$ between $1<z<1.5$. Specifically, we use $N(z) = (1+\tanh((z-1)/0.05))/2$.
\end{itemize}
In (iii) and (iv), boundary conditions at the new domain top boundary $z=1.5$ were no-slip.

\begin{figure*}[htb]
    \centering
    \includegraphics[scale=.83]{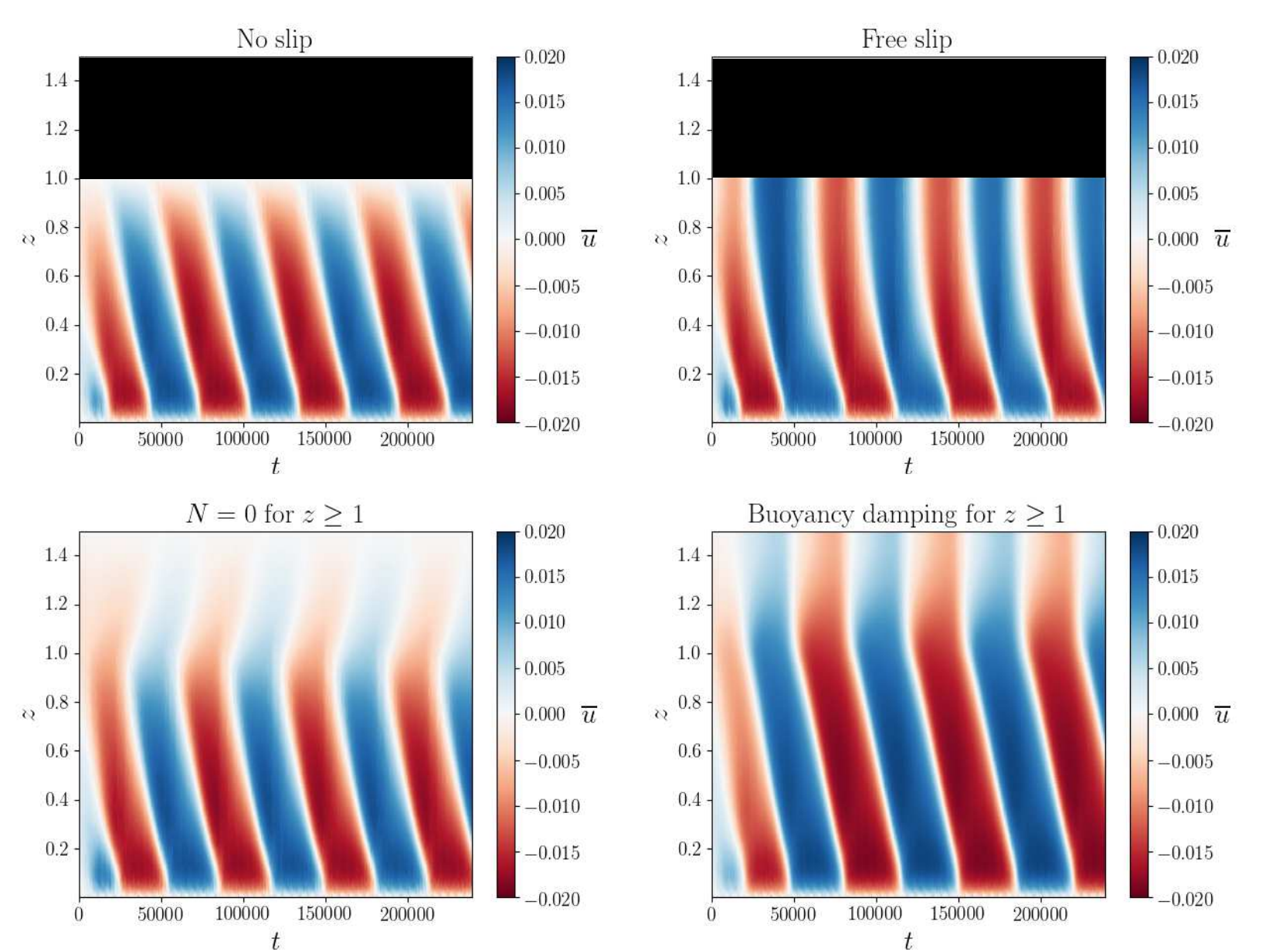}
    \caption{Hovm\"oller diagrams of the horizontal mean-flow driven by a multi-wave forcing for different top boundary conditions in our 2D numerical model. The forcing spectrum is gaussian centred at $\omega_0 = 0.2$ with standard deviation $\sigma = 10^{-1}$ and total energy $E_{tot} = 2\times 10^{-6}$. This corresponds to the wide forcing spectrum case shown in figure 1 of main text. In the top panels, the computational domain goes to $z=1$.}
    \label{fig:tbc_1}
\end{figure*}
\begin{figure*}[htb]
    \centering
    \includegraphics[scale=.83]{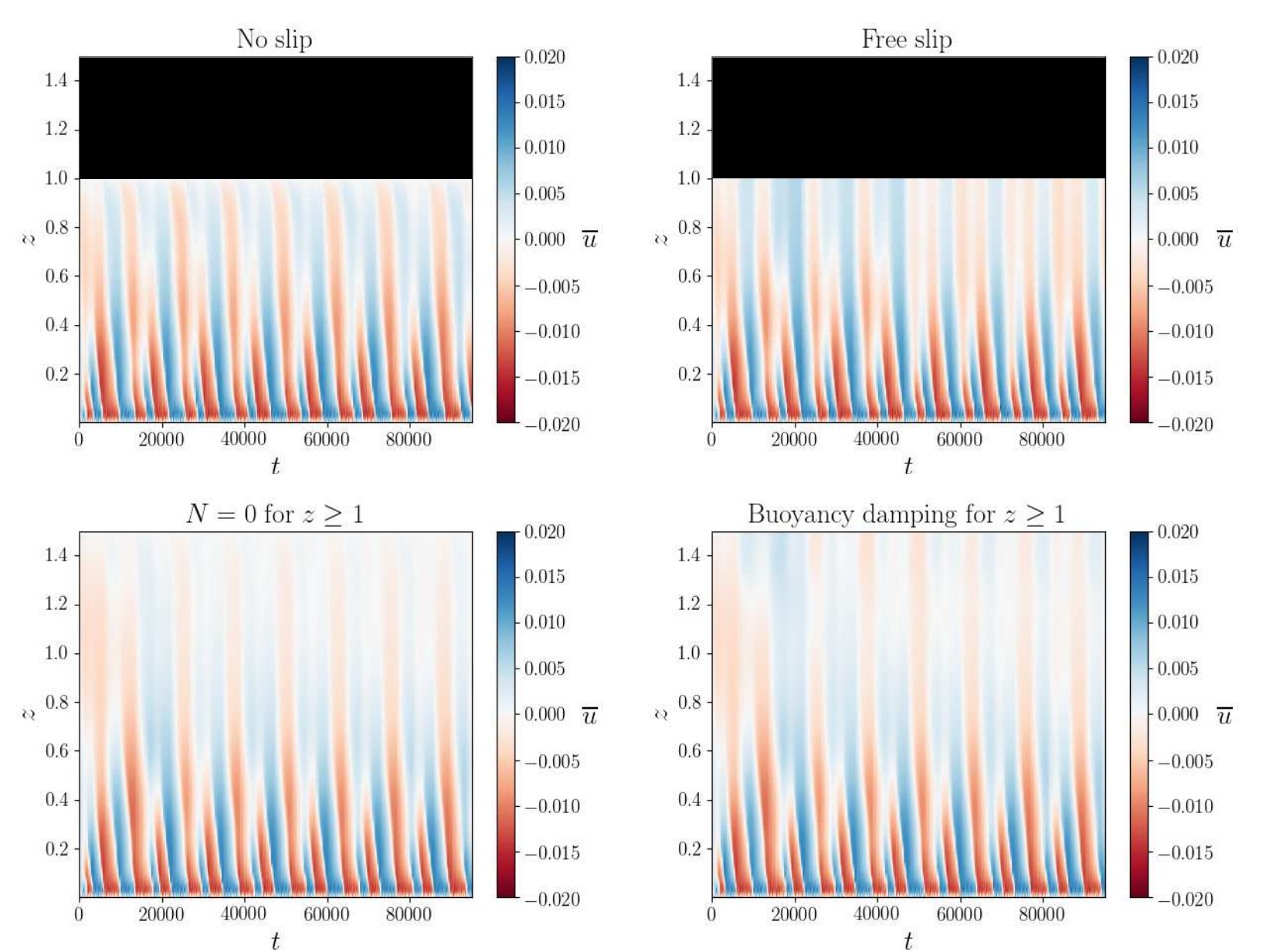}
    \caption{Same as Figure 3, but with a Gaussian forcing spectrum with standard deviation $\sigma = 2.8\times10^{-2}$.}
    \label{fig:tbc_2}
\end{figure*}

In figures \ref{fig:tbc_1} and \ref{fig:tbc_2} we plot Hovm\"{o}ller diagrams for simulations with two different values of $\sigma$.
In both cases, we find that the top boundary condition does not strongly affect the mean-flow evolution.
We find the same dynamical regimes as in the 1D model.
Therefore, using a 1D model with an easy to implement top boundary condition, such as free-slip, does not alter the results.
The free-slip condition for $\overline{u}$ that we use in our 1D simulations should be physically similar to having a layer with $N=0$.
A layer with $N=0$ is model for the mesosphere, which is weakly stratified ($N^2\sim 0$) compared to the stratosphere ($N^2 \neq 0$).
While wave reflection occurs at the interface, we find that in our 1D simulations, at most $\sim 10\%$ of the total wave energy reaches the top domain and is lost. This study of top boundary conditions suggests this does not qualitatively affect the results.

\clearpage
\subsection{Domain height}
In the main text we show that as the forcing spectrum becomes broader, at fixed energy, the mean-flow transitions from quasiperiodic to periodic.
This could be due to the excitation of high-frequency waves whose attenuation length is much larger than the domain height.
Using the forcing spectrum parameters of Figure 1 of the main text, we do find that the transition occurs at larger and larger $\sigma$ as the domain size increases.
For those parameters, the periodic mean-flow oscillations appear to be dictated by waves with large attenuation lengths.
However, this is not always the case.
In Figure \ref{fig:mu015}, we present simulations with $\omega_0=0.15$ and $E_{tot}=5\times 10^{-6}$.
For these parameters, we find the transition from quasiperiodic to periodic oscillations occurs at $\sigma \sim 2.5\times10^{-2}$.
This is lower than the transition value in Figure 1 of the main text, where we used the larger value of $\omega_0=0.2$.
In this case, all forced waves have attenuation lengths smaller than the domain height.
Thus, the top boundary is not required for periodic mean-flow oscillations at large $\sigma$.

\begin{figure}[h!]
    \centering
    \includegraphics[scale=.6]{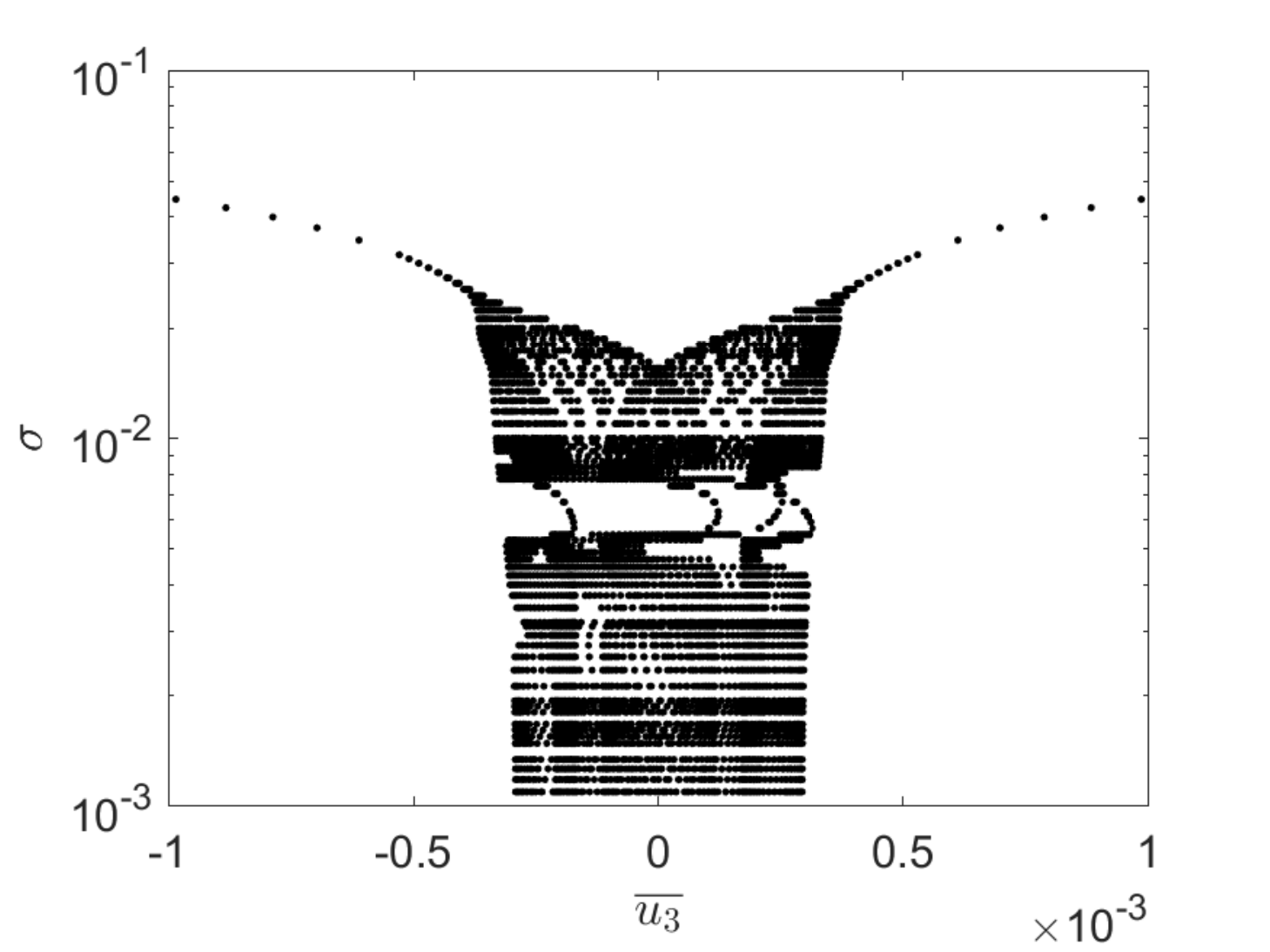}
    \caption{Poincar\'e map for simulations with different $\sigma$, with $\omega_0=0.15$ and $E_{tot} = 5\times 10^{-6}$. The transition from quasiperiodic to periodic reversals occurs at a value of $\sigma$ where all forced waves have an attenuation length smaller than domain height.}
    \label{fig:mu015}
\end{figure}

\end{widetext}

\end{document}